\documentclass[aps,prl,showpacs,superscriptaddress,prbib, twocolumn]{revtex4}
\usepackage{graphicx}
\usepackage{float}
\newcommand{\Fig}[1]{Fig.~\ref{#1}}

\begin{document}

\author{Lishan Zhao}
\affiliation{Scottish Universities Physics Alliance (SUPA), School of Physics and Astronomy, University of St.  Andrews, St. Andrews KY16 9SS, United Kingdom}
\affiliation{Max Planck Institute for Chemical Physics of Solids, N\"{o}thnitzer Str.~40, 01187 Dresden, Germany}
\author{Edward A. Yelland}
\affiliation{Scottish Universities Physics Alliance (SUPA), School of Physics and Astronomy, University of St.  Andrews, St. Andrews KY16 9SS, United Kingdom}
\affiliation{SUPA, School of Physics and Astronomy, and Centre for Science at Extreme Conditions, University of Edinburgh, Mayfield Road, Edinburgh EH9 3JZ, United Kingdom}
\author{Jan A. N. Bruin}
\affiliation{Scottish Universities Physics Alliance (SUPA), School of Physics and Astronomy, University of St.  Andrews, St. Andrews KY16 9SS, United Kingdom}
\affiliation{Max Planck Institute for Solid State Research, Heisenbergstr. 1, 70569 Stuttgart, Germany}
\author{Ilya Sheikin}
\affiliation{Laboratoire National des Champs Magn\'{e}tiques Intenses (LNCMI-EMFL), CNRS, UJF, F-38042 Grenoble, France}
\author{Paul C. Canfield}
\affiliation{Ames Laboratory and Department of Physics, Iowa State University, Ames, Iowa 50011, U.S.A.}
\author{Veronika Fritsch}
\affiliation{Experimental Physics VI, Center for Electronic Correlations and Magnetism, Institute of Physics, University of Augsburg, 86135 Augsburg, Germany}
\affiliation{Physikalisches Institut, Karlsruhe Institute of Technology, 76131 Karlsruhe, Germany}
\author{Hideaki Sakai}
\affiliation{Scottish Universities Physics Alliance (SUPA), School of Physics and Astronomy, University of St.  Andrews, St. Andrews KY16 9SS, United Kingdom}
\affiliation{Department of Physics, Osaka University, Toyonaka, Osaka 560-0043, Japan}
\author{Andrew P. Mackenzie}
\affiliation{Scottish Universities Physics Alliance (SUPA), School of Physics and Astronomy, University of St.  Andrews, St. Andrews KY16 9SS, United Kingdom}
\affiliation{Max Planck Institute for Chemical Physics of Solids, N\"{o}thnitzer Str.~40, 01187 Dresden, Germany}
\author{Clifford W. Hicks} 
\affiliation{Max Planck Institute for Chemical Physics of Solids, N\"{o}thnitzer Str.~40, 01187 Dresden, Germany}

\title{\large \textbf{Field-Temperature Phase Diagram and Entropy Landscape of CeAuSb$_2$}}

\date{17 Feb 2016}

\begin{abstract}

We report a field-temperature phase diagram and an entropy map for the heavy fermion compound CeAuSb$_2$. CeAuSb$_2$ orders antiferromagnetically below $T_N=6.6$~K, and has two metamagnetic
transitions, at 2.8 and 5.6~T. The locations of the critical endpoints of the metamagnetic transitions, which may play a strong role in the putative quantum criticality of CeAuSb$_2$ and related
compounds, are identified. The entropy map reveals an apparent entropy balance with Fermi liquid behavior, implying that above the N\'{e}el transition the Ce moments are incorporated into the Fermi
liquid. High-field data showing that the magnetic behavior is remarkably anisotropic are also reported.

\end{abstract}

\maketitle

\noindent \textbf{Introduction}
\\

CeAuSb$_2$ is a heavy-fermion system with the tetragonal $P4/nmm$ structure, moderate electrical anisotropy, and strong magnetic anisotropy.~\cite{Thamizhavel03} Although it has not been widely
studied, it shows strong phenomenological similarities with other cerium-based compounds that have received intense interest. A major theme of study of Ce-based systems is to understand and
tune the balance between Kondo and RKKY interaction: RKKY interaction couples localized spins and favors a magnetically ordered ground state, while strong Kondo interaction quenches local spins,
incorporating their entropy into the Fermi liquid. Data presented in this paper, and comparison with other compounds, suggest that CeAuSb$_2$ is on the border, with the effects of both Kondo and RKKY
interactions apparent in its bulk properties, but neither dominating.

\begin{figure}[t]
\includegraphics[width=85mm]{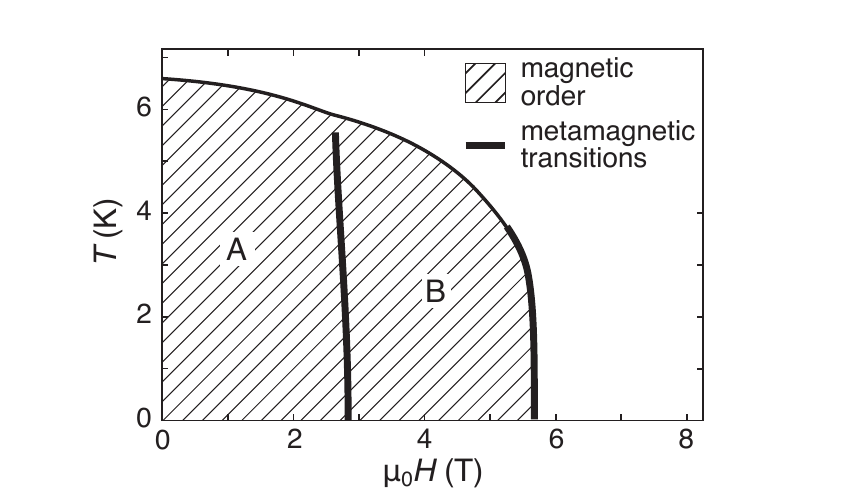}
\caption{\label{overviewPhaseDiagram} Field-temperature phase diagram of CeAuSb$_2$, with the magnetically ordered region and the metamagnetic transitions shown. The indicated phase boundaries are
from this work.}
\end{figure}

A field-temperature phase diagram of CeAuSb$_2$ is shown in \Fig{overviewPhaseDiagram}; the indicated phase boundaries are from this work, however many of its basic features were published in
Refs.~\cite{Balicas05} and~\cite{Lorenzer13}. At zero field, there is a N\'{e}el transition at $T_N = 6.6$~K. As the field is increased, there are two first-order metamagnetic transitions, at 2.8 and
5.6~T. The magnetic order terminates at the second metamagnetic transition.

There appear to be many materials with qualitatively similar field-temperature phase diagrams, including CeNiGe$_3$,~\cite{Mun10, Pikul03} CeRh$_2$Si$_2$,~\cite{Knafo10, Morales15} and
YbNiSi$_3$,~\cite{Budko07, Avila04} and the CeRu$_2$Si$_2$-based compounds (Ce$_{0.8}$La$_{0.2}$)Ru$_2$Si$_2$,~\cite{Mignot90, Aoki11, Shimizu12} CeRu$_2$(Si$_{0.9}$Ge$_{0.1}$)$_2$,~\cite{Mignot91,
Sugi08} and Ce(Ru$_{0.092}$Rh$_{0.08}$)$_2$Si$_2$.~\cite{Sekine92, Aoki12} All of these compouds show antiferromagnetism and, when the field is oriented along the easy axis, two metamagnetic
transitions. With the exception of Ce(Ru$_{0.92}$Rh$_{0.08}$)$_2$Si$_2$, the magnetic order of each terminates at the second metamagnetic transition. 

At low temperatures, all of the Ce-based compounds listed above have strongly anisotropic easy-axis magnetic susceptibilities. $\chi_c / \chi_a$ of CeAuSb$_2$ is 17 just above
$T_N$.~\cite{Thamizhavel03} $\chi_c / \chi_a$ of CeRh$_2$Si$_2$ is 5 just above its N\'{e}el temperature, and of CeRu$_2$Si$_2$, 15 at 10~K.~\cite{Settai97, Haen87} CeNiGe$_3$ is an orthorhombic
system where the $a$ axis is the easy axis; $\chi_a / \chi_b$ and $\chi_a / \chi_c$ are 11 and 17, respectively, just above its $T_N$.~\cite{Mun10}

Therefore, study of CeAuSb$_2$ is likely to have bearing on a range of other compounds. Comparison with the CeRu$_2$Si$_2$-based compounds is of particular interest.  CeRu$_2$Si$_2$ itself has a Kondo
temperature $T_K$ of $\sim 24$~K,~\cite{Haen87} and strong antiferromagnetic fluctuations at low temperature, but no static order;~\cite{Raymond98} the Kondo effect appears to win out over the RKKY
interaction by a small margin. Substitution can alter the balance: partial substitution of La for Ce, \textit{e.g.} decreases $T_K$ and induces the static antiferromagnetic order mentioned
above.~\cite{Amato89, Knafo09, Shimizu12} The substitution applies an effective negative pressure: the lattice is expanded, and when it is compressed again with (positive) hydrostatic pressure the
antiferromagnetism is suppressed.~\cite{Haen96} The similarity between the phase diagrams of CeAuSb$_2$ and substituted CeRu$_2$Si$_2$-based compounds suggests that CeAuSb$_2$ acts, broadly, as a
negative-pressure version of CeRu$_2$Si$_2$. Usefully, it is a version without intrinsic substitution disorder or, with reference to Ce$_{1-x}$La$_x$Ru$_2$Si$_2$, dilution of Ce spins.  It may allow
RKKY-Kondo crossover to be studied with positive rather than negative pressure.

The main aim of the present work is to refine the phase diagram of CeAuSb$_2$. The metamagnetic transitions
are thought to be first-order, but clear hysteresis has not been seen and their critical endpoints have not
been precisely located.~\cite{Balicas05} As will be elaborated upon in the Discussion, the endpoints may prove
crucial to possible quantum criticality in CeAuSb$_2$, and related compounds.  In addition to locating the
critical endpoints of CeAuSb$_2$, we also report an entropy map across the field-temperature phase diagram,
which yields both similarities and notable contrasts with the above compounds.  \\ \\

\noindent \textbf{Crystal growth}
\\

Single CeAuSb$_2$ crystals were grown by a self-flux method, similar to that described in
Refs.~\cite{Canfield92} and~\cite{Canfield01}. High purity ingots of Ce (99.99\%, Ames Laboratory), Au
(99.999\%, Alfa Aesar), and Sb (99.999\%, Alfa Aesar) were placed in an alumina crucible with a Ce:Au:Sb
atomic ratio of 1:6:12. The crucible was then sealed in an evacuated quartz ampoule and heated to
$1100^\circ$C for 10 hours, followed by cooling to $700^\circ$ over a period of 100 hours. The excess flux was
decanted with a centrifuge at $700^\circ$C. Measurement by energy-dispersive X-ray spectroscopy (EDX)
confirmed that the crystals are stoichiometric to within the 5\% measurement precision. The residual
resistivity ratios of the crystals used in this work were between 6 and 9. A photograph of an as-grown
CeAuSb$_2$ crystal is shown in Fig.~2.  
\\

\noindent \textbf{Results: resistivity}
\\

For measurement of resistivity, samples were cut into narrow bars with a wire saw. The as-grown samples were
naturally thin along the $c$ axis, so polishing to reduce thickness was not necessary. The samples were
measured with a typical four-terminal method using a lock-in amplifier, with typically a 100~$\mu$A excitation
current at a frequency on the order of 100~Hz. Contacts to the sample were made with DuPont 6838 silver paste, baked
at 180$^\circ$~C for 2.5 hours. 

\begin{figure}[ptb]
\includegraphics[width=85mm]{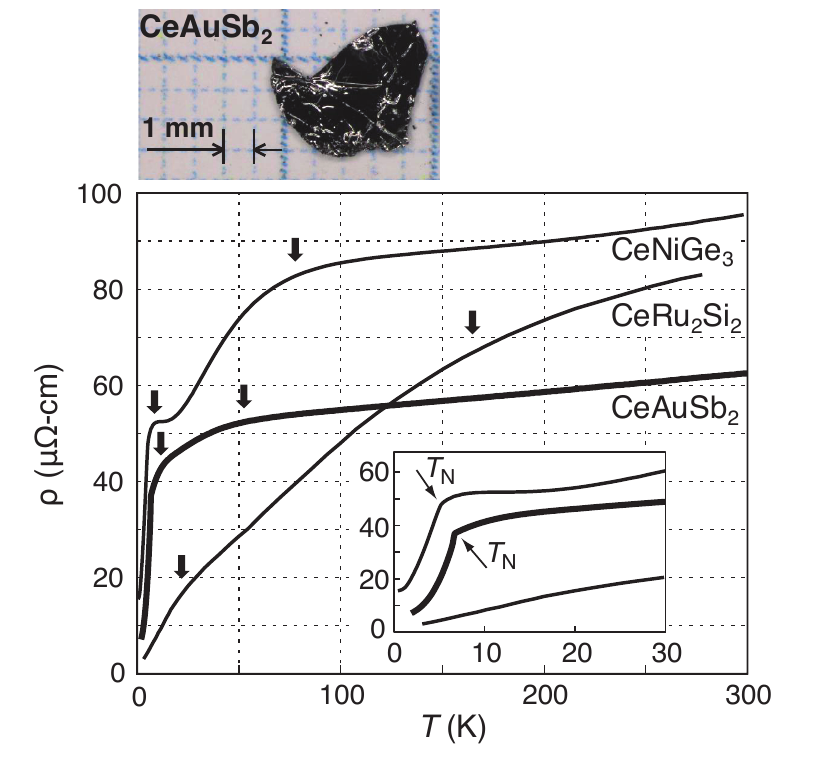}
\caption{\label{resistivityComparison} Top: a photograph of an as-grown crystal of CeAuSb$_2$. Bottom: comparison of $\rho(T)$ of CeNiGe$_3$ (Ref.~\cite{Mun10}), CeRu$_2$Si$_2$ (Ref.~\cite{Haen87}),
and CeAuSb$_2$ (this work). For CeRu$_2$Si$_2$ and CeAuSb$_2$, the current $\mathbf{I}$ is in the plane, and for CeNiGe$_3$, an orthorhombic material, $\mathbf{I} \parallel \mathbf{\hat{c}}$. The arrows mark
broad shoulders, discussed in the text.} 
\end{figure}

We start in \Fig{resistivityComparison} with a comparison of the resistivities $\rho(T)$ of CeAuSb$_2$, CeRu$_2$Si$_2$, and CeNiGe$_3$. The room-temperature values are similar; respective comparison
of CeAuSb$_2$ with LaAuSb$_2$~\cite{Seo12} and CeRu$_2$Si$_2$ with LaRu$_2$Si$_2$~\cite{Haen87} show that scattering from cerium spins accounts for roughly half the resistivity at room-temperature,
and a greater portion as $T$ is reduced. The resistivities also all show two broad shoulders, marked by the arrows in \Fig{resistivityComparison}. The higher-temperature shoulders are due to thermal
occupation of excited crystal electric field states. The origin of the lower-temperature shoulders may differ from compound to compound; in CeRu$_2$Si$_2$ it is attributed to the Kondo
effect.~\cite{Haen87} In both CeNiGe$_3$ and CeAuSb$_2$ the shoulder is at a lower temperature than in CeRu$_2$Si$_2$; both these compounds also show N\'{e}el order, so the lower-temperature shoulder
could be due to onset of short-range magnetic order at $T > T_N$, or the Kondo effect, or a combination. 

\begin{figure}[ptb]
\includegraphics[width=85mm]{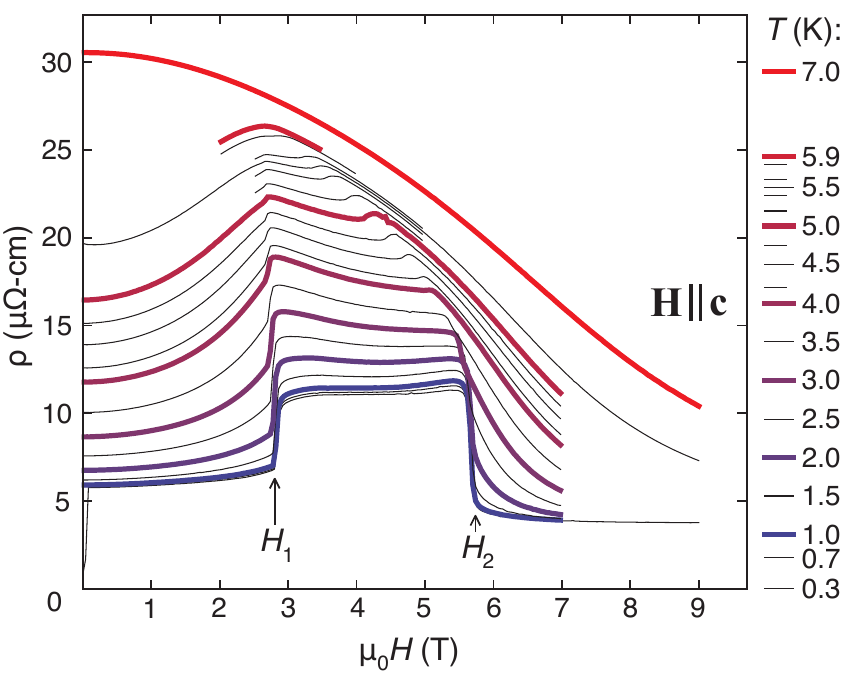}
\caption{\label{resistivity}In-plane resistivity of CeAuSb$_2$ against field, applied along the $c$ axis. The
plotted data are from increasing-field ramps.}
\end{figure}

Low-temperature measurements of the resistivity of CeAuSb$_2$ were done on two samples from the same growth batch, with cross sections $\approx$$180 \times 130$ and $\approx$$250 \times 150$~$\mu$m
(the shorter dimension along the $c$ axis). They were measured together in an adiabatic demagnetization refrigerator. \Fig{resistivity} shows the resistivity of one sample against field at various
fixed temperatures. At the lowest temperatures, $\rho$ increases sharply at the first metamagnetic transition (at applied field $H_1$), and decreases sharply at the second ($H_2$). 
Elevated resistivity between $H_1$ and $H_2$ is also seen in CeNiGe$_3$~\cite{Mun10}, YbNiSi$_3$~\cite{Budko07}, and CeRh$_2$Si$_2$.~\cite{Knafo10}

\begin{figure}[ptb]
\includegraphics[width=85mm]{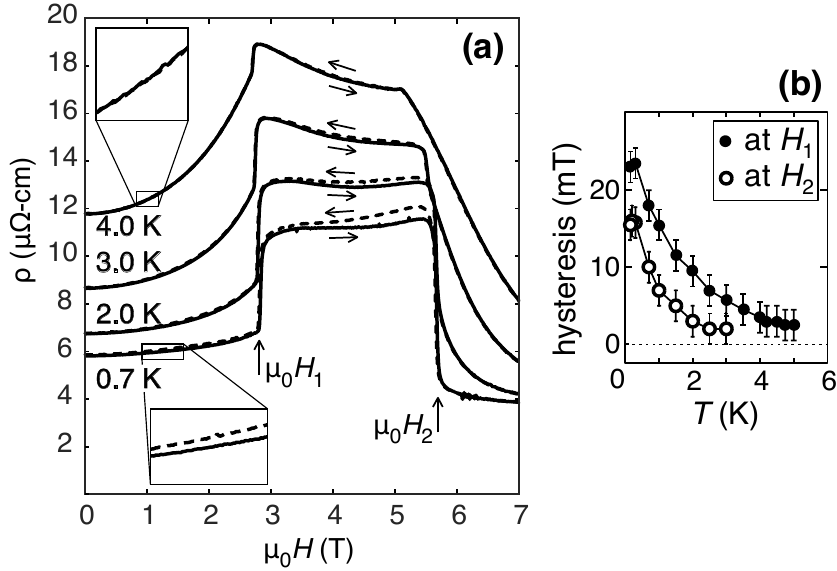}
\caption{\label{hysteresis}(a) Data from increasing- and decreasing-field ramps, plotted together.
(b) Difference in the transition fields between the increasing- and decreasing-field ramps, plotted
against temperature. For all temperatures, the ramp rate was 0.1~T/min.}
\end{figure}

Data from increasing-field and decreasing-field ramps are shown together in \Fig{hysteresis}; magnet hysteresis is excluded by measuring the field with a Hall sensor placed near the sample.
Hysteresis in the metamagnetic transition fields shows that the transitions are first-order. The magnitude of the hysteresis against temperature is shown in panel~(b): it decreases steadily as the
temperature is increased (proving that it is not an artefact of instrument hysteresis), disappearing within experimental resolution above $\sim$5~K for the first and above $\sim$3~K for the second
transition.

\begin{figure}[ptb]
\includegraphics[width=85mm]{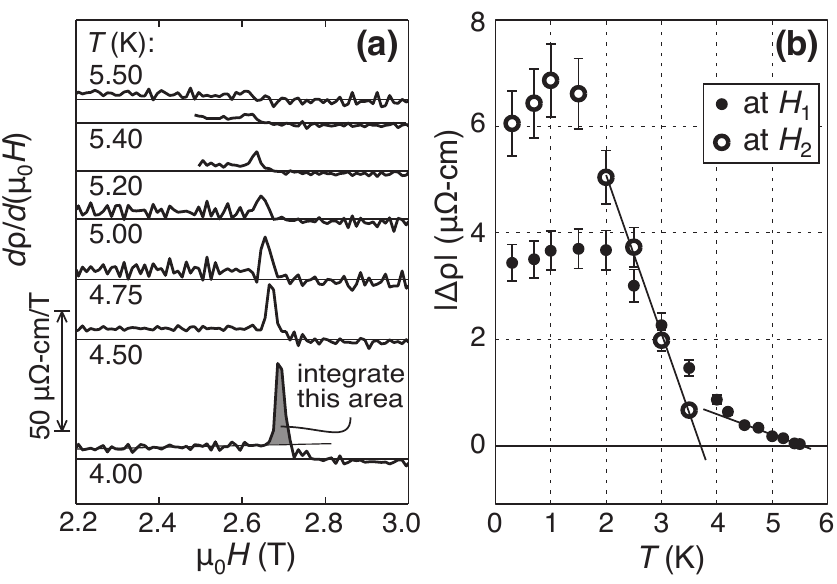}
\caption{\label{firstOrderJumps} (a) $d \rho / d (\mu_0 H)$ against $\mu_0 H$ at the first transition,
at temperatures approaching the critical endpoint. On the bottom curve, the method for determining the
resistivity jump $\Delta \rho$ is illustrated: a background is fit to the data to the left (right) of the peak
for the first (second) metamagnetic transition, and the peak integrated. (b) $|\Delta \rho|$ against
temperature.}
\end{figure}

Rounding of the transitions and remnant instrument hysteresis make it impossible to pinpoint the temperatures where the hysteresis ends.  To locate the endpoints more precisely, we analysed the
magnitudes of the resistivity jumps, $|\Delta \rho|$, determined by integrating the peaks in the derivative $d\rho / d(\mu_0 H)$ as illustrated in \Fig{firstOrderJumps}(a). Rounding introduces
moderate systematic error, but by following the same procedure at each temperature random error is minimized.  $|\Delta \rho|$ against $T$ is shown in \Fig{firstOrderJumps}(b); the error bars are the
estimated systematic error. Linear fits through the higher-temperature points locate the endpoints of the first and second transitions at 5.6 and 3.7~K, respectively. 

In \Fig{hysteresis}, hysteresis is apparent not only in the transition fields, but also in the magnitude of
the resistivity across the magnetically ordered region: $\rho$ is larger on decreasing- than increasing-field
ramps. Hysteresis between $H_1$ and $H_2$ has been reported before;~\cite{Balicas05,Lorenzer13} its origin is
not resolved, although antiferromagnetic domain walls are a natural possibility. We add two further
observations: (1) There is hysteresis below $H_1$, as well as between $H_1$ and $H_2$. (2) The magnitude of
the hysteresis decreases approximately linearly as the temperature is increased, disappearing between 4 and
5~K.  
\\

\noindent \textbf{Results: specific heat capacity}
\\

The specific heat capacity $C$ of a sample roughly 2~mm across and 0.1~mm thick, with mass 4.0~mg, was
measured in a Quantum Design Physical Properties Measurement System. The relaxation time method was used: at each
point the sample temperature was raised by 2\%, then the relaxation time was measured.  

\begin{figure}[ptb]
\includegraphics[width=85mm]{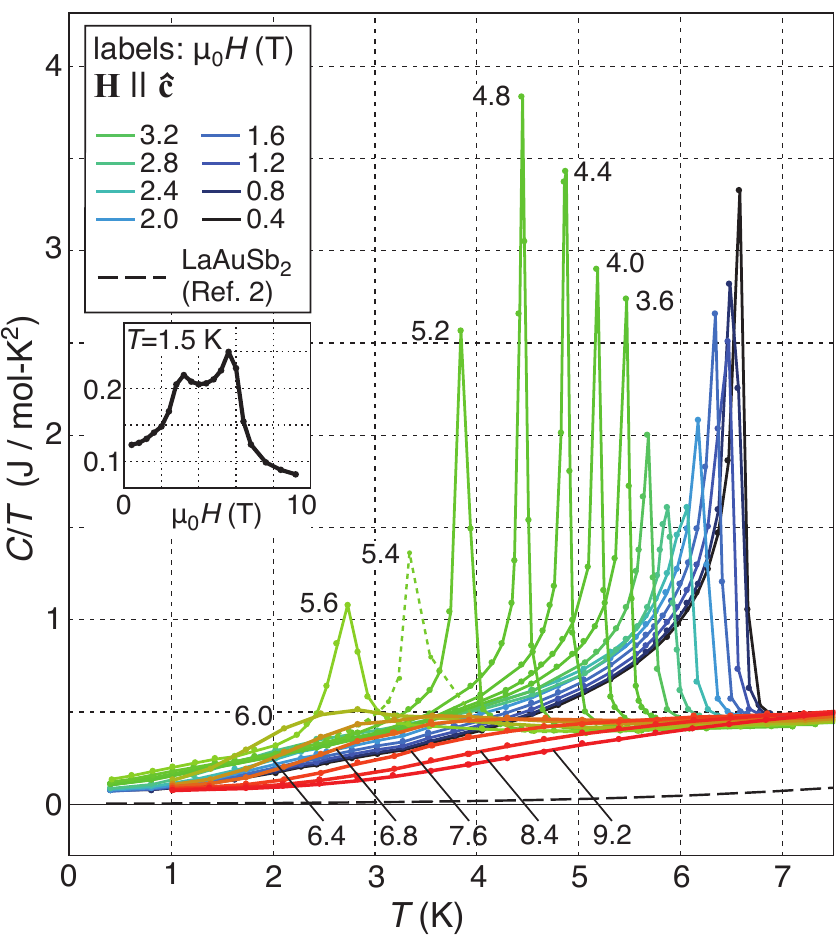}
\caption{\label{specificHeat}Main panel: $C/T$ of CeAuSb$_2$ at selected fields. Inset: $C/T$ against field at $T=1.5$~K.}
\end{figure}

$C/T$ against temperature at selected magnetic fields is plotted in \Fig{specificHeat}. $C/T$ of LaAuSb$_2$, from Ref.~\cite{Balicas05}, is also shown, as an estimate of the nonmagnetic contribution;
up to at least $\sim$7~K it is much smaller than $C/T$ of CeAuSb$_2$. The dominant feature in the low-temperature heat capacity of CeAuSb$_2$ is the N\'{e}el transition. At low fields, the peak at
$T_N$ is relatively narrow. It is broader in the vicinity of the first critical endpoint, at 2.6~T and 5.6~K, and becomes very sharp as the field is increased towards the second critical endpoint, at
5.2~T and 3.7~K.

\begin{figure}[ptb]
\includegraphics[width=85mm]{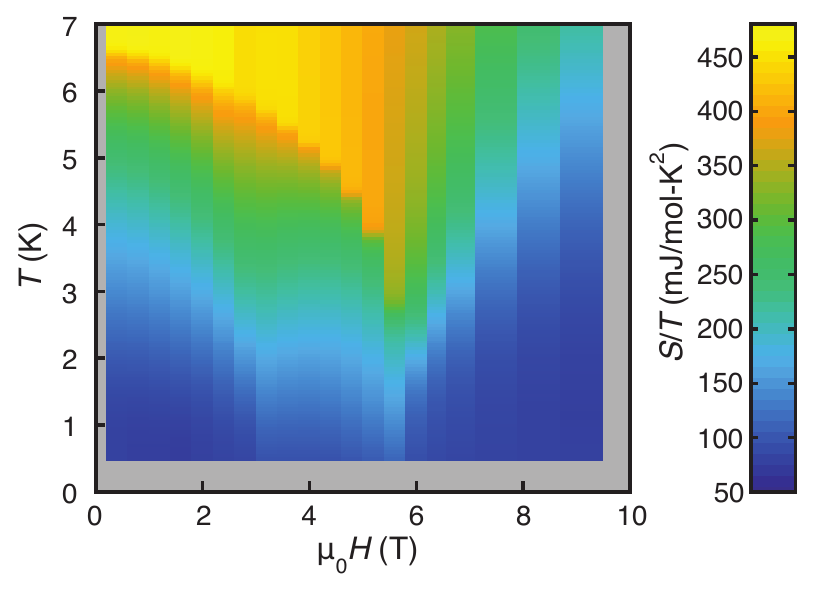}
\caption{\label{entropy}Entropy over temperature of CeAuSb$_2$ against temperature and
$c$-axis magnetic field.} 
\end{figure}

The inset in \Fig{specificHeat} shows $C/T$ against field at $T=1.5$~K. It is higher between $H_1$ and $H_2$ than on either side. This behavior has also been established for CeNiGe$_3$~\cite{Mun10},
Ce(Ru$_{0.92}$Rh$_{0.08}$)$_2$Si$_2$~\cite{Aoki12}, CeRh$_2$Si$_2$~\cite{Morales15}, and (though less pronounced) Ce$_{0.9}$La$_{0.1}$Ru$_2$Si$_2$~\cite{Aoki11}.

$C/T$ may be integrated from 0~K at each field to yield the entropy, $S$. To do so, an extrapolation to 0~K is required, though the data extend to low enough temperature that the precise
form of the extrapolation is not critical. We take, at each field, a linear extrapolation from the lowest-temperature data point to 79.5~mJ/mol-K$^2$ at 0~K, which is an apparent base value in the
data. A map of the resulting entropy, divided by temperature, is shown in \Fig{entropy}. Below the N\'{e}el transition, the entropy is generally higher over the field range $H_1<H<H_2$ than on either
side.

At lower fields, $S/T$ and $C/T$ closely match above $T_N$: at $\mu_0 H=0.4$~T and $T=7$~K, $S/T$ and $C/T$ are 0.47 and 0.48~J/mol-K$^2$, respectively. In other words, the magnetic order maintains
entropy balance with a Fermi-liquid-like, $T$-independent $C/T$. (Subtracting $C/T$ of LaAuSb$_2$, as an estimate for the phonon contribution, makes little difference: $S/T$ and $C/T$ are respectively
revised to 0.45 and 0.41~J/mol-K$^2$.) Entropy balance with a Fermi liquid suggests that the Ce $4f$ moments are fully incorporated into the Fermi liquid below some temperature that exceeds $T_N$,
such that in the absence of magnetic order $C/T$ would be $T$-independent down to 0~K. This behavior is in contrast to CeNiGe$_3$, YbNiSi$_3$, and Ce$_{0.7}$La$_{0.3}$Ru$_2$Si$_2$. Based on analysis
of published data,~\cite{Mun10, Budko07, Besnus87} and taking reasonable extrapolations of $C/T$ to 0~K, we find that in these compounds $S/T$ at 7~K (above $T_N$ for each) exceeds $C/T$ by more than
a factor of two. (Specifically, $S/T$ and $C/T$ at 7~K are respectively 0.65 and 0.30~J/mol-K$^2$ for CeNiGe$_3$, $\approx$0.63 and 0.27~J/mol-K$^2$ for YbNiSi$_3$, and $\approx$0.58 and
0.21~J/mol-K$^2$ for Ce$_{0.7}$La$_{0.3}$Ru$_2$Si$_2$.) In these compounds the $4f$ moments appear to give a quasi-independent contribution to $S/T$ that is in addition to a Fermi liquid contribution:
it appears that the moments are strongly disordered by $T \sim 7$~K, and so make a large contribution to the total entropy, but at 7~K the dominant contribution to $C/T$ is from the Fermi liquid.

It is also notable that above $T_N$, $C/T$ of CeAuSb$_2$ drops very quickly to its Fermi liquid value, whereas
$C/T$ of each of CeNiGe$_3$, YbNiSi$_3$, and Ce$_{0.7}$La$_{0.3}$Ru$_2$Si$_2$ has a strong decaying tail that
extends at least a few K above $T_N$. Such tails are common in local-moment systems, \textit{e.g.} in
PdCrO$_2$,~\cite{Takatsu09} and arise from gradual onset of short-range magnetic order above $T_N$. They
provide further evidence that in these compounds the $4f$ moments and Fermi liquid are quasi-independent
systems, while in CeAuSb$_2$ they are not.

At low fields, the entropy of CeAuSb$_2$ above $T_N$ is a substantial fraction of $R \log 2$: at 0.4~T and 7~K, $S$ = \mbox{3.3~J/mol-K} = $0.57 R\log 2$. The $T>T_N$ entropy is gradually suppressed
as the field is increased; it appears that the heavy fermion state is gradually suppressed through polarization of the Ce moments.
\\

\noindent \textbf{Results: high-field measurements}
\\

\begin{figure}[t]
\includegraphics[width=85mm]{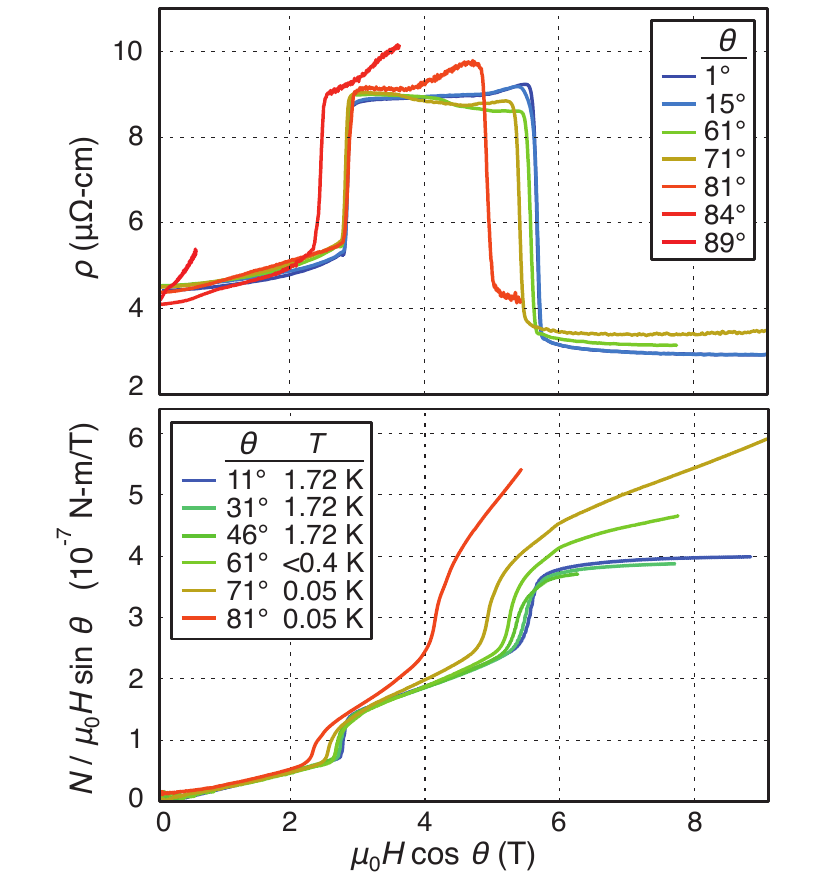}
\caption{\label{angleStudy}Measurements of CeAuSb$_2$ in fields up to 35~T. $\theta$ is the angle between
$\mathbf{H}$ and the $c$ axis. Upper panel: resistivity against $c$-axis field $\mu_0 H \cos(\theta)$. Lower
panel: torque $N$, divided by the in-plane component of the field, against $\mu_0 H \cos(\theta)$.
Measurements at 1.72~K were performed in St Andrews, and at lower temperatures in the high field laboratory in
Grenoble.}
\end{figure}

To probe the magnetic anisotropy, resistivity and torque magnetometry measurements up to 35~T were performed at the Laboratoire National des Champs Magn\'{e}tiques Intenses in Grenoble, France. The
samples were mounted on a rotatable platform, to vary the field angle.  The transport sample was a bar with cross-sectional area $230 \times 90$~$\mu$m. The dimensions of the torque
magnetometry sample were $\sim 250 \times 250 \times 50$~$\mu$m. Results are shown in \Fig{angleStudy}.

Previously published measurements showed that $T_N$ is almost independent of in-plane field up to 18~T.~\cite{Balicas05} Our measurements show similarly that the metamagnetic transition fields are
remarkably unaffected by the presence of a strong in-plane field. In the figure, $\rho$ is plotted against the $c$-axis field, $H_{\mathbf{\hat{c}}} = H \cos \theta$, with $\theta$ the angle between
$\mathbf{H}$ and the $c$ axis.  The form of $\rho(H_{\mathbf{\hat{c}}})$ changes little as the in-plane field is increased, even to the point that $\mathbf{H}$ is only a few degrees out of the plane. 

The metamagnetic transitions are also apparent in the torque data. Plotted in the lower panel of \Fig{angleStudy} is the torque $N$ divided by the in-plane field $H \sin \theta$: if the field-induced
magnetization is pinned to the $c$ axis, and is a function of $H_\mathbf{\hat{c}}$ alone (\textit{i.e.} independent of in-plane field), then the graph of $N/H \sin \theta$ against $H_\mathbf{\hat{c}}$
will be independent of field angle. The data show that this is essentially the case for CeAuSb$_2$ up to $\theta \sim 70^\circ$, confirming that the magnetism of CeAuSb$_2$ is strongly easy-axis-type.
As the field gets very close to the $ab$ plane, the metamagnetic transitions move to lower $H_\mathbf{\hat{c}}$.  
\\

\noindent \textbf{Phase diagram and Discussion}
\\

\begin{figure}[t]
\includegraphics[width=85mm]{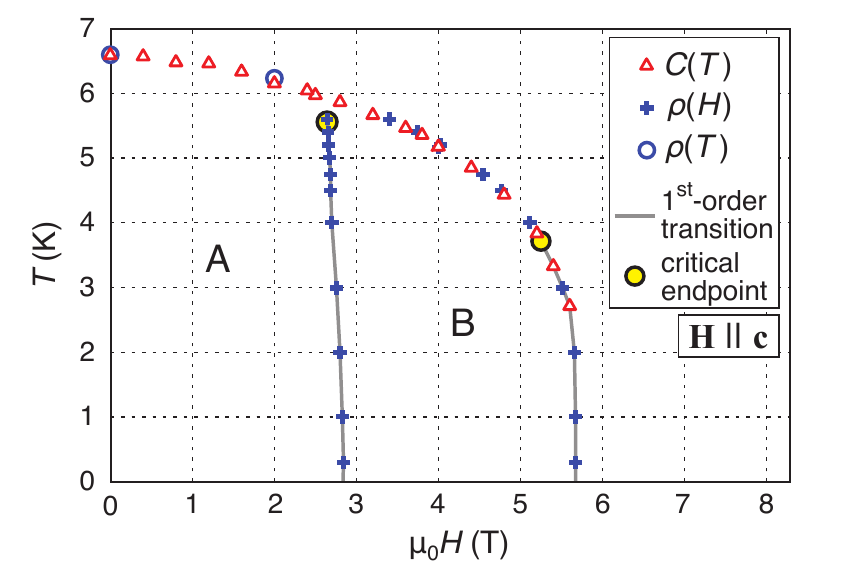}
\caption{\label{phaseDiagram}Phase diagram of CeAuSb$_2$, derived from measurements of $\rho$ against $H$,
$\rho$ against $T$, and specific heat against $T$.}
\end{figure}

\Fig{phaseDiagram} shows the field-temperature phase diagram derived from the resistivity and specific heat data presented above. The first-order metamagnetic transition lines, and their critical
endpoints, are indicated. The first metamagnetic line separates regions of magnetic order that may be designated the A and B phases. The line slopes leftward as $T$ is raised, consistent with the
observation that the B phase has higher entropy than the A phase. In contrast to a previous report, we do not find evidence for an intermediate phase along either metamagnetic line.~\cite{Lorenzer13} 

The first critical endpoint, at a temperature of 5.6~K, appears to lie $\approx$0.4~K below $T_N$ at that field, although the present data cannot exclude with high confidence that it is not in fact on
the N\'{e}el line. If it is indeed below $T_N$, it would be interesting to determine whether the boundary between the A and B phases continues to $T_N$ as a crossover or a second-order transition; the
former implies adiabatic continuity between the two phases, and the latter an additional symmetry breaking.

Quantum criticality is a major theme in study of heavy-fermion systems, and in studies of criticality the
locations of critical endpoints is vital knowledge. For example, many key properties of CeRu$_2$Si$_2$ are
explainable by a missed quantum critical endpoint at $\mu_0 H \approx 7.6$~T.~\cite{Weickert10} At the famous
field-driven antiferromagnetic quantum critical point of YbRh$_2$Si$_2$, it is proposed that it is in fact
close proximity of a metamagnetic critical endpoint and an antiferromagnetic QCP, rather than the
antiferromagnetic QCP alone, that drives the observed divergences in the Sommerfeld coefficient and magnetic
susceptibility.~\cite{Misawa08, Hackl11, Gegenwart05} The superconductivity of URhGe~\cite{Yelland11} and the
anomalous phase in Sr$_3$Ru$_2$O$_7$~\cite{Grigera04} both form around low-temperature metamagnetic critical
endpoints.

In CeAuSb$_2$, the second critical endpoint is at a temperature of 3.7~K.  This is more than half the maximum $T_N$, so quantum critical scaling to $T \sim 0$~K is not expected.  However it would be a
very compelling experiment to track the endpoints with pressure, and to attempt to drive them to 0~K.  The antiferromagnetic order of both CeRh$_2$Si$_2$ and CeNiGe$_3$ can be suppressed with
pressure, with superconductivity appearing in a window of pressure around the antiferromagnetic QCP.~\cite{Kotegawa06, Araki02} It would be interesting to determine whether metamagnetic quantum
criticality is also involved in this superconductivity. Regarding CeAuSb$_2$, a published pressure study at $H=0$ showed that pressure initially increases the temperature of the first resistivity
shoulder and decreases $T_N$.~\cite{Seo12}

As described above, entropy balance with a Fermi liquid suggests that Kondo coupling is important in CeAuSb$_2$, incorporating the $4f$ moments into the Fermi liquid by some temperature that exceeds
$T_N$; strong Kondo coupling probably onsets at the temperature of the first resistivity shoulder, at $\sim$12~K. Phenomenologically, CeAuSb$_2$ therefore appears to be intermediate to
CeRu$_2$Si$_2$, where Kondo coupling is strong enough that static magnetic order never emerges, and, \textit{e.g.}, CeNiGe$_3$ and Ce$_{0.7}$La$_{0.3}$Ru$_2$Si$_2$, where the moments and conduction
electrons appear to be quasi-decoupled at all temperatures. 

We also note a strong similarity between the phenomenology of CeAuSb$_2$ and Sr$_3$Ru$_2$O$_7$, which was in fact the original motivation for this study of CeAuSb$_2$: both show strongly enhanced
resistivity over a finite window of field (in the case of Sr$_3$Ru$_2$O$_7$, between $\approx$7.9 and 8.1~T), bounded by first-order metamagnetic transitions, and over which the entropy is also
higher.~\cite{Grigera04, Rost09} Sr$_3$Ru$_2$O$_7$ is very clearly an itinerant system, while in CeAuSb$_2$ the moments probably have strong local character below $T_N$. This similar behavior in spite
of this substantial difference suggests a deep link between the two systems.
\\

\noindent \textbf{Conclusion}
\\

In conclusion, we have produced a refined field-temperature phase diagram of CeAuSb$_2$, and an entropy map spanning the region of magnetic order. We have also highlighted similarities between
CeAuSb$_2$ and other compounds. The observed metamagnetic transitions were sufficiently sharp to resolve clear hysteresis, and to locate their critical endpoints, showing that CeAuSb$_2$ can now be
grown with sufficiently low disorder to make it a useful reference material and target for further study. 

We acknowledge useful discussion with Manuel Brando and Christoph Geibel. We thank Hanoh Lee for advice on crystal growth. We thank the EPSRC and the Max Planck Society for financial support. We also
acknowledge the support of the LNCMI-CNRS, member of the European Magnetic Field Laboratory (EMFL). E.A.Y. acknowledges support from the Royal Society. P.C.C. was supported, in part, by the U.S.
Department of Energy, Office of Basic Energy Science, Division of Materials Sciences and Engineering through the Ames Laboratory. Ames Laboratory is operated for the U.S. Department of Energy by Iowa
State University under Contract No. DE-AC02-07CH11358. V.F. acknowledges support by the Deutsche Forschungsgemeinschaft through FOR 960. H.S. gratefully acknowledges fellowships from the Canon
Foundation.

The raw data for the figures in this article can be found in Supplemental Material.


\begin{thebibliography}{99}

\bibitem{Thamizhavel03}A. Thamizhavel, T. Takeuchi, T. Okubo, M. Yamada, R. Asai, S. Kirita, A. Galatanu, E.
Yamamoto, T. Ebihara, Y. Inada, R. Settai, and Y. Onuki. Anisotropic electrical and magnetic properties of
Ce$T$Sb$_2$ ($T$=Cu, au, and Ni) single crystals. \textit{Phys. Rev. B} \textbf{68} 054427 (2003).

\bibitem{Balicas05}L. Balicas, S. Nakatsuji, H. Lee, P. Schlottmann, T.P. Murphy, and Z. Fisk. Magnetic
field-tuned quantum critical point in CeAuSb$_2$. \textit{Phys. Rev. B} \textbf{72} 064422 (2005).

\bibitem{Lorenzer13}K.-A. Lorenzer, A.M. Strydom, A. Thamizhavel, and S. Paschen. Temperature-field phase
diagram of quantum critical CeAuSb$_2$. \textit{Phys. Status Solidi B} \textbf{250} 464 (2013).

\bibitem{Mun10}E.D. Mun, S.L. Bud'ko, A. Kreyssig, and P.C. Canfield. Tuning low-temperature physical
properties of CeNiGe$_3$ by magnetic field. \textit{Phys. Rev. B} \textbf{82} 054424 (2010).

\bibitem{Pikul03}A.P. Pikul, D. Kaczorowski, T. Plackowski, A. Czopnik, H. Michor, E. Bauer, G. Hilscher, P.
Rogl, and Yu. Grin. Kondo behavior in antiferromagnetic CeNiGe$_3$. \textit{Phys. Rev. B} \textbf{67} 224417
(2003).

\bibitem{Knafo10}W. Knafo, D. Aoki, D. Vignolles, B. Vignolle, Y. Klein, C. Jaudet, A. Villaume, C. Proust,
and J. Flouquet. High-field metamagnetism in the antiferromagnet CeRh$_2$Si$_2$. \textit{Phys. Rev. B}
\textbf{81} 094403 (2010).

\bibitem{Morales15}A. Palacio Morales, A. Pourret, G. Seyfarth, M.T. Suzuki, D. Braithwaite, G. Knebel, D.
Aoki, and J. Flouquet. Fermi surface instabilities in CeRh$_2$Si$_2$ at high magnetic field and pressure.
\textit{Phys. Rev. B} \textbf{91} 245129 (2015).

\bibitem{Budko07}S.L. Bud'ko, P.C. Canfield, M.A. Avila, and T. Takabatake. Magnetic-field tuning of the
low-temperature state of YbNiSi$_3$. \textit{Phys. Rev. B} \textbf{75} 094433 (2007).

\bibitem{Avila04}M.A. Avila, M. Sera, and T. Takabatake. YbNiSi$_3$: An antiferromagnetic Kondo lattice with
strong exchange interaction. \textit{Phys. Rev. B} \textbf{70} 100409R (2004).

\bibitem{Mignot90}J.-M. Mignot, J.-L. Jacoud, L.-P. Regnault, J. Rossat-Mignod, P. Haen, P. Lejay, Ph.
Boutrouille, B. Hennion, and D. Petitgrand. Neutron diffraction study of (Ce,La)Ru$_2$Si$_2$ alloys in an
external field. \textit{Physica B} \textbf{163} 611 (1990).

\bibitem{Aoki11}D. Aoki, C. Paulsen, T.D. Matsuda, L. Malone, G. Knebel, P. Haen, P. Lejay, R. Settai, Y.
$\overline{\mbox{O}}$nuki, and J. Flouquet. Pressure Evolution of the Magnetic Field Induced Ferromagnetic
Fluctuations through the Pseudo-Metamagnetism of CeRu$_2$Si$_2$. \textit{J. Phys. Soc. Japan} \textbf{80}
053702 (2011).

\bibitem{Shimizu12}Y. Shimizu, Y. Matsumoto, K. Aoki, N. Kimura, H. Aoki. Anomalous Transport Properties via
the Competition between the RKKY Interaction and the Kondo Effect in Ce$_x$La$_{1-x}$Ru$_2$Si$_2$. \textit{J.
Phys. Soc. Japan} \textbf{81} 044707 (2012).

\bibitem{Mignot91}J.M. Mignot, L.P. Regnault, J.L. Jacoud, J. Rossat-Mignod, P. Haen, and P. Lejay.
Incommensurabilities and metamagnetism in the heavy-fermion alloys (Ce$_{0.8}$La$_{0.1}$)Ru$_2$Si$_2$ and
CeRu$_2$(Si$_{0.9}$Ge$_{0.1}$)$_2$. \textit{Physica B} \textbf{171} 357 (1991).

\bibitem{Sugi08}M. Sugi, Y. Matsumoto, N. Kimura, T. Komatsubara, H. Aoki, T. Terashima, and S. Uji. Fermi
Surface Properties of CeRu$_2$(Si$_{1-x}$Ge$_x$)$_2$ in Magnetic Fields above the Metamagnetic Transitions.
\textit{Phys. Rev. Lett.} \textbf{101} 056401 (2008).

\bibitem{Sekine92}C. Sekine, T. Sakakibara, H. Amitsuka, Y. Miyako, and T. Goto. Magnetic Properties and Phase
Diagram of Ce(Ru$_{1-x}$Rh$_x$)$_2$Si$_2$ ($0 \leq x < 0.5$). \textit{J. Phys. Soc. Japan} \textbf{61} 4536
(1992).

\bibitem{Aoki12}D. Aoki, C. Paulsen, H. Kotegawa, F. Hardy, C. Meingast, P. Haen, M. Boukahil, W. Knafo, E.
Ressouche, S. Raymond, and J. Flouquet. Decoupling between Field-Instabilities of Antiferromagnetism and
Psuedo-Metamagnetism in Rh-Doped CeRu$_2$Si$_2$ Kondo Lattice. \textit{J. Phys. Soc. Japan} \textbf{81} 034711
(2012).

\bibitem{Settai97}R. Settai, A. Misawa, S. Araki, M. Kosaki, K. Sugiyama, T. Takeuchi, K. Kindo, Y. Haga, E.
Yamamoto, and Y. $\bar{\mbox{O}}$nuki. Single Crystal Growth and Magnetic Properties of CeRh$_2$Si$_2$.
\textit{J. Phys. Soc. Japan} \textbf{66} 2260 (1997).

\bibitem{Haen87}P. Haen, J. Flouquet, F. Lapierre, P. Lejay, and G. Remenyi. Metamagnetic-like Transition in
CeRu$_2$Si$_2$? \textit{J. Low Temp. Phys.} \textbf{67} 391 (1987).

\bibitem{Raymond98}S. Raymond, L.P. Regnault, S. Kambe, J. Flouquet., and P. Lejay. Switching of the magnetic
interactions from antiferromagnetic to ferromagnetic in the heavy-fermion compound CeRu$_2$Si$_2$ under high
magnetic field. \textit{J. Phys.: Condensed Matter} \textbf{10} 2363 (1998).

\bibitem{Amato89}A. Amato, D. Jaccard, J. Sierro, P. Haen, P. Lejay, and J. Flouquet. Transport Properties
under Magnetic Fields of the Heavy Fermion System CeRu$_2$Si$_2$ and Related Compounds (Ce,La)Ru$_2$Si$_2$.
\textit{J. Low Temp. Phys.} \textbf{77} 195 (1989).

\bibitem{Knafo09}W. Knafo, S. Raymond, P. Lejay, and J. Flouquet. Antiferromagnetic criticality at a
heavy-fermion quantum phase transition. \textit{Nat. Phys.} \textbf{5} 753 (2009).

\bibitem{Haen96}P. Haen, F. Lapierre, J. Voiron, J. Flouquet. Vanishing of magnetic order in
Ce$_{0.8}$La$_{0.2}$Ru$_2$Si$_2$ under pressure. \textit{J. Phys. Soc. Japan} \textbf{65} (Suppl. B) 27
(1996).

\bibitem{Canfield92}P.C. Canfield and Z. Fisk. Growth of single-crystals from metallic fluxes.
\textit{Philosophical Magazine B} \textbf{65} 1117 (1992).

\bibitem{Canfield01}P.C. Canfield and I.R. Fisher. High-temperature solution growth of intermetallic single
crystals and quasicrystals. \textit{J. Crystal Growth} \textbf{225} 155 (2001).

\bibitem{Seo12}S. Seo, V.A. Sidorov, H. Lee, D. Jang, Z. Fisk, J.D. Thompson, and T. Park. Pressure effects on
the heavy-fermion antiferromagnet CeAuSb$_2$. \textit{Phys. Rev. B} \textbf{85} 205145 (2012).

\bibitem{Besnus87}M.J. Besnus, P. Lehmann, and A. Meyer. Heat capacity study of the (La-Ce)Ru$_2$Si$_2$ and
(Ce-Y)Ru$_2$Si$_2$ Kondo systems. \textit{Journal of  Magnetism and Magnetic Materials} \textbf{63 \& 64} 323
(1987).

\bibitem{Takatsu09}H. Takatsu, H. Yoshizawa, S. Yonezawa, Y. Maeno. Critical behavior of the metallic
triangular-lattice Heisenberg antiferromagnet PdCrO$_2$. \textit{Phys. Rev. B} \textbf{79} 104424 (2009).

\bibitem{Weickert10}F. Weickert, M. Brando, F. Steglich, P. Gegenwart, and M. Garst. Universal signatures of
the metamagnetic quantum critical endpoint: Application to CeRu$_2$Si$_2$. \textit{Phys. Rev. B} \textbf{81}
134438 (2010).

\bibitem{Misawa08}T. Misawa, Y. Yamaji, and M. Imada. YbRh$_2$Si$_2$: Quantum Tricritical Behavior in
Itinerant Electron System. \textit{J. Phys. Soc. Japan} \textbf{77} 093712 (2008).

\bibitem{Hackl11}A. Hackl and M. Vojta. Zeeman-Driven Lifshitz Transition: A Model for the Experimentally
Observed Fermi-Surface Reconstruction in YbRh$_2$Si$_2$. \textit{Phys. Rev. Lett.} \textbf{106} 137002 (2011).

\bibitem{Gegenwart05}P. Gegenwart, J. Custers, Y. Tokiwa, C. Geibel, and F. Steglich. Ferromagnetic quantum
critical fluctuations in YbRh$_2$(Si$_{0.95}$Ge$_{0.05}$)$_2$. \textit{Phys. Rev. Lett.} \textbf{94} 076402
(2005).

\bibitem{Yelland11}E.A. Yelland, J.M. Barraclough, W. Wang, K.V. Kamenev and A.D. Huxley. High-field
superconductivity at an electronic topological transition in URhGe. \textit{Nat. Physics} \textbf{7} 890
(2011).

\bibitem{Grigera04}S.A. Grigera, P. Gegenwart, R.A. Borzi, F. Weickert, A.J. Schofield, R.S. Perry, T. Tayama,
T. Sakakibara, Y. Maeno, A.G. Green, A.P. Mackenzie. Disorder-Sensitive Phase Formation Linked to Metamagnetic
Quantum Criticality. \textit{Science} \textbf{306} 1154 (2004). 

\bibitem{Kotegawa06}H. Kotegawa, K. Takeda, T. Miyoshi, S. Fukushima, H. Hidaka, T.C. Kobayashi, T. Akazawa,
Y. Ohishi, M. Nakashima, A. Thamizhavel, R. Settai, and Y. $\bar{\mbox{O}}$nuki. Pressure-induced
superconductivity emerging from antiferromagnetic phase in CeNiGe$_3$. \textit{J. Phys. Soc. Japan}
\textbf{75} 044713 (2006).

\bibitem{Araki02}S. Araki, M. Nakashima, R. Settai, T.C. Kobayashi, Y. $\bar{\mbox{O}}$nuki. Pressure-induced
superconductivity in an antiferromagnet CeRh$_2$Si$_2$. \textit{J. Phys.: Condensed Matter} \textbf{14} L377
(2002).

\bibitem{Rost09}A.W. Rost, R.S. Perry, J.-F. Mercure, A.P. Mackenzie, S.A. Grigera. Entropy Landscape of Phase
Formation Associated with Quantum Criticality in Sr$_3$Ru$_2$O$_7$. \textit{Science} \textbf{325} 1360 (2009).

\end{thebibliography}
\end{document}